\documentclass[reprint,prl,showpacs,superscriptaddress]{revtex4-2}
\usepackage{amssymb}
\usepackage{amsmath}    
\usepackage{graphicx}   
\usepackage{verbatim}

\usepackage{color}      
\usepackage{booktabs} 

\begin{document}
\title{Radiation from polarized vacuum in a laser-particle collision}
\author{M. Jirka}
\affiliation{The Extreme Light Infrastructure ERIC, ELI Beamlines Facility, Za Radnici 835, 25241 Dolni Brezany, 
Czech 
Republic}
\affiliation{Faculty of Nuclear Sciences and Physical Engineering, Czech
Technical University in Prague, Brehova 7, 115 19 Prague, Czech Republic}

\author{P. Sasorov}
\affiliation{Extreme Light Infrastructure ERIC, Za Radnici 835, 25241 Dolni Brezany, Czech Republic}

\author{S. V. Bulanov}
\affiliation{Extreme Light Infrastructure ERIC, Za Radnici 835, 25241 Dolni Brezany, Czech Republic}
\affiliation{National Institutes for Quantum and Radiological Science and 
Technology (QST), Kansai Photon Science Institute, 8-1-7 Umemidai, Kizugawa, 
Kyoto 619–0215, Japan}

\begin{abstract}
The probability of photon emission of a charged particle 
traversing 
a strong 
field becomes modified if vacuum polarization is considered.
This feature is important for fundamental quantum electrodynamics processes 
present in extreme astrophysical environments and can be studied in a collision of a charged particle with a strong 
laser field.
We show that for today's available 700~GeV (6.5~TeV) protons and 
the 
field provided 
by the next generation of lasers, the emission spectra 
peak is enhanced due to vacuum polarization effect
by 30\% (suppressed by 65\%) in comparison to the traditionally considered 
Compton 
process. 
This striking phenomenon offers a novel path to the laboratory-based 
manifestation of vacuum polarization.
\end{abstract}

\maketitle

According to quantum theory, under the action of a strong field, the vacuum 
behaves as a medium with a modified refractive index for a probe 
electromagnetic 
(EM) 
radiation
\cite{Heisenberg1936,Beresteckij2008,Toll1952}.
As the next generation of lasers will provide very strong EM fields, they will 
enable the study of the fundamental physics of photon-photon interactions 
\cite{Mourou2006,Marklund2006,DiPiazza2012,Gonoskov2022,Fedotov2023}.
Several setups for measuring the signatures of vacuum polarization on probe 
photons using the intense lasers have been already suggested (see 
Refs.~\cite{King2016,Karbstein2020} and references therein).
The basic idea is to detect probe photons whose kinematics or polarization 
properties differ from the original ones \cite{Karbstein2019}.
However, the signature of vacuum polarization on probe photons still has 
not been observed \cite{Karbstein2022}. 
The possibility of experimentally studying the quantum vacuum is 
attractive also for modeling the processes important for astrophysics as 
the intense lasers will allow 
us to reach the strong-field regime of quantum electrodynamics (QED), similar 
to that thought to exist in extreme environments where the vacuum polarization effects are expected \cite{Piran2005}.
For example, the emission of magnetars is thought to be polarized due the 
vacuum polarization effect \cite{Heyl2002,Mignani2016} (see also the remarks in \cite{Capparelli2017}).
Here we demonstrate a novel feature: the radiation from a charged particle 
traversing a strong field region provided e.g.  by magnetars or by the 
ultra intense laser can be drastically enhanced or reduced 
once 
vacuum polarization is taken into account.

When a strong EM field is applied to a vacuum, the virtual 
electron-positron 
pairs become polarized.
For a photon traversing a strong field region, the situation can be 
conveniently described by assigning an index of refraction to the vacuum that 
allows for various non-linear phenomena such as birefringence and 
vacuum Cherenkov radiation \cite{Erber1966}.
Thus, as a consequence of vacuum polarization, the probability of  photon 
emission by a charged particle interacting with a strong 
EM
field is governed by the synergy between synchrotron-like and Cherenkov 
mechanisms 
\cite{Erber1976,Schwinger1976,Rynne1978}.
It is important to note that due to the synergism, it is an indecomposable
physical process when the resulting radiation exhibits features which 
cannot be 
recovered by superposing the constituents \cite{Schwinger1976,Erber1976}.
As the most intense EM fields are nowadays produced 
only by lasers \cite{Yoon2021}, there is growing interest in observing the 
synergic 
Compton-Cherenkov 
(SCC) radiation \cite{Bulanov2019} 
of charged particles in a strong EM field.
The counterplay between the Cherenkov radiation and non-linear Compton 
scattering has been studied in \cite{Dremin2002,Bulanov2019,Sasorov2022} and the possibility of Cherenkov radiation in the cosmic microwave background radiation has been discussed in \cite{Dremin2002astro}.
In comparison to these papers, our study considers the whole energy spectrum of emitted photons, not only the limit of low-energy photons, i.e. Cherenkov ones.
Here we show, that the SCC process exhibits so-far unexplored features in 
photon 
emission distribution.
This striking phenomenon offers a novel path to the laboratory-based manifestation 
of vacuum polarization.

In this paper, we explore the features in photon spectra emerging due to 
the emission of a charged particle in a 
polarized vacuum and demonstrate differences from the traditionally considered tree-level, i.e. lowest-order, process of
non-linear Compton scattering in which vacuum 
polarization effects are not considered \cite{Ritus1985}.
We show that sizeable differences in photon emission spectra are expected for 
protons 
of today's
available energy and the EM field provided 
by the next generation of lasers.
Namely, the amount of 
emitted energy can be either significantly enhanced or reduced as a consequence 
of synergic 
nature of photon emission.
These features of photon emission could serve as a 
signature 
of vacuum polarization.
While it is well known that in non-linear Compton 
scattering the 
recoil 
effect reduces the emitted energy \cite{Blackburn2020}, we show that recoil 
also controls the amount 
of energy (not)emitted 
due to the effects of vacuum polarization implemented in SCC 
process.
In particular, a recoil experienced by the emitting particle, that has been 
considered negligible in the case of quantum Cherenkov radiation 
\cite{Ginzburg1996}, 
plays 
an important role in the above-mentioned more general processes as it suppresses 
the effect of vacuum polarization on photon 
emission.

The interaction of the charged particle and photon with a strong 
EM field 
is characterized by two Lorentz invariant parameters $ 
\chi=e\hbar\sqrt{-\left( 
F^{\mu\nu}p_{\nu}\right)^{2} }/m^{3}c^{4} $ and $ \chi_\gamma= 
\hbar\sqrt{-\left( F^{\mu\nu}k_{\nu}\right)^{2}}/m_{e}cE_{\mathrm{S}}$, 
respectively, where $ 
e $ is  the 
elementary charge, $ \hbar $ is the reduced Planck constant, $ m $ is the 
particle mass, $
F_{\mu\nu}=\partial_{\mu}A_{\nu}-\partial_{\nu}A_{\mu}$ 
is the EM field tensor, $ A_{\nu} $ is the four-potential, $ 
p_{\nu} $ is the 
four-momentum of the emitting particle, $ k_{\nu} $ is the photon four-wave 
vector, $ 
m_{e} $ denotes the electron mass and $ c $ is the speed of light in a vacuum
\cite{Ritus1985}.
The Schwinger-Sauter field $E_{\mathrm{S}} 
=m_{e}^{2}c^{3}/e\hbar$ is approximately of $1.33\times10^{16}~\mathrm{V/cm} 
$
\cite{Schwinger1951}.
The strength of the EM wave is characterized by the Lorentz 
invariant 
parameter $a_{0}=eE_{0}/m\omega_{0} c $, where $ E_{0} $ is the amplitude of 
the electric field, $ \omega_{0}=2\pi c/\lambda $ is the angular frequency and 
$ \lambda $ is the wavelength.

Since we are interested in photon emission in the quantum vacuum characterized 
by an index of refraction $ n=1+\Delta n $, where $ \lvert\Delta 
n\rvert \ll 1 $ is the change of refraction index due to vacuum polarization, 
we need to derive the 
formula for photon emission that includes the effect of the index of 
refraction.
Therefore we use the semiclassical approach, where the motion of the energetic 
particle is 
treated classically while photon emission is treated quantum mechanically, 
as it presents a powerful method for numerical calculation of photon emission 
\cite{Baier1998}.
For $ a_{0}\gg1 $ (i.e. constant crossed field approximation) and under the weak-field approximation, 
i.e. $ 
\chi^{2}\gg \mathcal{F}, 
\mathcal{G} $ and $ \mathcal{F},\mathcal{G}\ll1 $ 
where $ \mathcal{F}=\lvert 
\mathbf{E}^{2}-\mathbf{B}^{2}\rvert e^{2}\hbar^{2}/m^{4}c^{6} $ and $ 
\mathcal{G}= \lvert 
\mathbf{E}\cdot\mathbf{B}\rvert e^{2}\hbar^{2}/m^{4}c^{6} $ are the 
normalized Poincaré invariants of EM field, the 
following approach is applicable to any particle-field configuration 
characterized by the same value of a parameter $ \chi $
\cite{Ritus1985,Beresteckij2008}.

The energy $ \mathcal{W} $ radiated by a charged particle per unit 
frequency interval $ d\omega $ 
and per unit solid angle $ d\Omega $ is \cite{Wistisen2015}
\begin{equation}\label{W}
\dfrac{d^{2}\mathcal{W}}{d\omega\, d\Omega}=\dfrac{e^{2}}{4\pi^{2}c}\left(\dfrac{\mathcal{E'}^{2}+\mathcal{E}^{2}}{2\mathcal{E}^{2}}\lvert \boldsymbol{I} \rvert^{2} + \dfrac{\hbar^{2}\omega^{2}m_{e}^{2}c^{4}}{2\mathcal{E}^{4}}\lvert J \rvert^{2} \right),
\end{equation}
where
\begin{equation}\label{I}
\boldsymbol{I}=\int_{-\infty}^{\infty}\dfrac{\boldsymbol{\nu}\times\left[\left( 
\boldsymbol{\nu}-\boldsymbol{\beta} \right)\times\dot{\boldsymbol{\beta}}  
\right] }{\left( 1-\boldsymbol{\nu}\cdot\boldsymbol{\beta} \right) 
^{2}}\exp\left[i\omega't\left( 1-\boldsymbol{\nu}\cdot\boldsymbol{\beta}\right) 
\right] \,dt,
\end{equation}
and
\begin{equation}\label{J}
J=\int_{-\infty}^{\infty}\dfrac{\boldsymbol{\nu}\cdot\dot{\boldsymbol{\beta}}}{\left(1-\boldsymbol{\nu}\cdot\boldsymbol{\beta}
 \right) ^{2}}\exp\left[i\omega't\left( 
1-\boldsymbol{\nu}\cdot\boldsymbol{\beta}\right) \right] \,dt.
\end{equation}
In Eqs.~\eqref{W}-\eqref{J} $ \mathcal{E} $ is the particle energy, $ 
\hbar\omega $ is the photon 
energy, $ 
\mathcal{E'}=\mathcal{E}-\hbar\omega $, $ 
\omega'=\omega\mathcal{E}/\mathcal{E'} $ and $ \boldsymbol{\nu} $ is the 
direction 
of emission \cite{Wistisen2015}.
Vector $ \boldsymbol{\beta}=\boldsymbol{v}/c $, $ \boldsymbol{v} $ is the particle velocity and $ \dot{\boldsymbol{\beta}} $ represents the differentiation of $ \boldsymbol{\beta} $ with respect to time $ t $.

The derivation procedure of Eq.~\eqref{W} is outlined in Ref.~\cite{Baier1998}.
It
considers the following relation
\begin{equation}\label{exponent}
\hat{\mathcal{E}}\left(\hat{p}-\hbar\hat{k}\right)
=\sqrt{\mathcal{E'}^{2}\underbrace{+2\hbar\omega\mathcal{E}
\left(1-\boldsymbol{\nu}\cdot\boldsymbol{\beta}\right)}_{A}
\underbrace{-\hbar^{2}(\omega^{2}-\boldsymbol{k}^{2}c^{2})}_{B}},
\end{equation}
where $ \hat{\mathcal{E}} $, $ \hat{p} $, $ 
\hat{k} $ are the operators of particle energy, particle 
momentum, and photon momentum, respectively (see Eq.~(2.20) in  
\cite{Baier1998}).
Vector $ \boldsymbol{k} $ is the wave vector of the emitted photon, in a medium 
$ \lvert \boldsymbol{k} \rvert=(1+\Delta n)\omega/c $ \cite{Baier1999}.
In our case, the role of a medium is played by polarized vacuum \cite{Ritus1970}.
However, the semiclassical approach can only be considered as an approximative solution in the regime where $\lvert \Delta n \rvert \ll 1$. For an exact solution, a full one-loop calculation of the Compton scattering process in a strong EM field needs to be performed as it would automatically include the effect of vacuum polarization.
We note that the properties of photons characterized by $k_{\nu}^{2}\neq0$ have been studied in Refs. \cite{Cronstrom1977,Becker1977,Heinzl2016}.

At first, we consider the case of $ \Delta 
n =0 $, that corresponds to the situation in which vacuum polarization 
effects are 
neglected.
Therefore, the term $B$ vanishes in Eq.~\eqref{exponent}.
As shown in 
Ref.~\cite{Baier1998}, the factor $ 1-\boldsymbol{\nu}\cdot\boldsymbol{\beta} $ 
appearing in part $A$ 
of Eq.~\eqref{exponent}
directly passes to the 
argument of the exponentials in Eqs.~\eqref{I} and \eqref{J}.
When an ultra-relativistic particle is assumed ($\gamma\gg 1$, $\gamma$ 
is the particle Lorentz factor), then $ 
1-\boldsymbol{\nu}\cdot\boldsymbol{\beta}\simeq1/(2\gamma^{2})+\theta^{2}/2+\mathcal{O}\left(
 1/\gamma^{4}\right)  
$ is considered in Eq.~\eqref{W}, where $ \theta\approx1/\gamma $ is an angle 
between $ \boldsymbol{\nu} $ and $ \boldsymbol{\beta} $ \cite{Baier1998}.
In such a case, the result of the integration of Eq.~\eqref{W} over $ \Omega $ 
can 
be recast in the form that is used for 
numerical 
calculation of photon emission power spectrum in non-linear Compton scattering 
\begin{equation}\label{W_Blackburn}
\dfrac{\mathrm{d} \mathcal{P}}{\mathrm{d} \omega}=\dfrac{\alpha 
\omega}{\sqrt{3}\pi\gamma^{2}}\left[
\dfrac{\zeta^{2}-2\zeta+1}{1-\zeta}K_{2/3}\left( 
\eta\right)-\int_{\eta}^{\infty}K_{1/3}\left( y\right) \mathrm{d}y   \right],
\end{equation}
where $ \alpha=e^{2}/\hbar c $ is the 
fine structure constant, $ \zeta=\hbar \omega/\mathcal{E}$, 
$ \eta=2\zeta/\left[3\chi\left( 1-\zeta\right)  \right] $, 
and $ K_{l}\left( m\right)  $ are the modified Bessel functions of the second 
kind 
\cite{Baier1998,Blackburn2020}.
Thus considering $ \omega^{2}-\boldsymbol{k}^{2}c^{2}=0 $ in Eq.~\eqref{exponent} refers to radiation  process in an external field \cite{Baier1998}.

On the other hand, the term  $ \omega^{2}-\boldsymbol{k}^{2}c^{2}\neq 0 $ is essential when radiation in a medium is considered \cite{Baier1999}.
In the case of a polarized vacuum, we assume that $ \lvert \boldsymbol{k} \rvert=(1+\Delta n)\omega/c $  holds within the interaction region and thus we need to consider $ \Delta n \neq 0 $ in Eq.~\eqref{exponent} \cite{Baier1999}.
Neglecting the small terms $ \propto \Delta n/\gamma^{4} $ and $ \propto 
\Delta n^{2} $ we obtain
\begin{equation}
1-\boldsymbol{\nu}\cdot\boldsymbol{\beta}\approx\dfrac{1}{2\gamma^{2}}+\dfrac{\theta^{2}}{2}-\Delta
 n,
\end{equation}
and
\begin{equation}
B\approx2\hbar^{2}\omega^{2}\Delta n.
\end{equation}
Since the term $ A $ in Eq.~\eqref{exponent} is approximately equal to
\begin{equation}\label{C}
A\approx2\hbar\omega\mathcal{E}\left( 
\dfrac{1}{2\gamma^{2}}+\dfrac{\theta^{2}}{2}-\Delta n\right),
\end{equation}
we obtain
\begin{equation}\label{CplusD}
A+B=2\hbar\omega\mathcal{E}\left[ 
\dfrac{1}{2\gamma^{2}}+\dfrac{\theta^{2}}{2}-\Delta n \left( 
1-\dfrac{\hbar\omega}{\mathcal{E}}\right) \right].
\end{equation}
The last term in the square brackets of Eq.~\eqref{CplusD} is a correction 
accounting for the effect of the index of refraction. 
This effect weakens as photon energy $\hbar\omega$ increases.
Including this correction into Eq.~\eqref{W} and following the procedure outlined in Ref.~\cite{Wistisen2015}, we obtain the differential power spectrum for photon emission in a quantum vacuum characterized by $ n=1+\Delta n $
\begin{widetext}
\begin{equation}\label{dP}
\dfrac{d^{2}\mathcal{P}}{d\omega\, 
d\theta}=e^{2}\omega^{2}\rho\left\lbrace 
\dfrac{\mathcal{E'}^{2}+\mathcal{E}^{2}}{2\mathcal{E'}^{2}}F\left[
\xi^{2}\mathrm{Ai}'^{2}(\xi)+\dfrac{\theta^{2}}{F}\mathrm{Ai}^{2}(\xi)\right]+\dfrac{\omega^{2}m^{2}}{2\mathcal{E'}^{2}\mathcal{E}^{2}}F
 \mathrm{Ai}^{2}\left(\xi \right) \right\rbrace,
\end{equation}
\end{widetext} 
where
\begin{equation}\label{F}
F=\dfrac{1}{\gamma^{2}}+\theta^{2}-2\Delta
 n 
\left(1-\dfrac{\hbar\omega}{\mathcal{E}}\right),
\end{equation}
\begin{equation}
\xi=\left(\dfrac{\omega'\rho}{2}\right)^{2/3}F,
\end{equation}
$\mathrm{Ai(\xi)}$ is the Airy function, and $ \rho $ is the radius 
of curvature of the particle trajectory.

Let us now consider a photon characterized by $ \chi_{\gamma} $ 
counterpropagating 
with respect 
to the constant crossed EM field 
($\boldsymbol{E}\perp\boldsymbol{B}$, $ \lvert \boldsymbol{E} \rvert = \lvert 
\boldsymbol{B} \rvert $).
Due to the possibility of creating an electron-positron pair by a photon 
in 
a 
field, the field can be considered as a homogeneous anisotropic medium with 
dispersion and absorption and thus can be characterized by assigning the 
complex index 
of refraction $ 
\tilde{n}(\chi_{\gamma})=1+\Delta \tilde{n}(\chi_{\gamma})$.
In the limits $ \chi_{\gamma}\ll1 $ and $ \chi_{\gamma}\gg1 $, the complex $ 
\Delta \tilde{n}(\chi_{\gamma}) $ can be expressed as
\begin{widetext}
\begin{equation}\label{DeltaN}
\Delta \tilde{n}\left( \chi_{\gamma}\right)  = \dfrac{\alpha 
m_{e}^{2}c^{4}}{2(\hbar\omega)^{2}}\times\begin{cases}
\left( \dfrac{11\mp 3}{90\pi}\chi_{\gamma}^{2} -i\sqrt{\dfrac{3}{2}}\dfrac{3\mp 
1}{16}\chi_{\gamma}\exp\left( -8/3\chi_{\gamma}\right)\right) 
&\text{, $\chi_{\gamma}\ll1$}\\
 -\dfrac{5\mp1}{28\pi^{2}}\sqrt{3}\Gamma^{4}\left(\dfrac{2}{3} 
\right) \left( 1-i\sqrt{3}\right) \left( 3\chi_{\gamma}\right) ^{\frac{2}{3}} 
&\text{, 
$\chi_{\gamma}\gg1$}
\end{cases}
\end{equation}
\end{widetext}
where the minus (plus) sign corresponds to the parallel (perpendicular) 
polarization of the photon with respect 
to the electric field, and $ \Gamma(x) $ is the Gamma function 
\cite{Ritus1970,Narozhny1969}.
For a given $ \chi_{\gamma} $, the corresponding value of $ \Delta 
\tilde{n}(\chi_{\gamma}) $ can be obtained from Eq.~(2.7) in 
Ref.~\cite{Ritus1970}.
The real part of the index of refraction determines the angle and intensity of 
SCC radiation, while its imaginary part describes photon absorption by 
the EM
field as it 
is connected with a probability of electron-positron pair 
production by the emitted photon that becomes important for large values of $\chi_{\gamma}$ \cite{Ritus1970,Ritus1985}.
For $\chi_{\gamma}\gg1$, the ratio of imaginary and real parts of $\tilde{n}(\chi_{\gamma}) $  in the studied parameter range is of the same order as if $\Re \left[ \Delta\tilde{n} \left( \chi_{\gamma}\right)\right]=0$ was considered.
As shown in Refs.~\cite{Beresteckij2008,Baier1998,Baier1999}, for $\Re \left[ \Delta\tilde{n} \left( \chi_{\gamma}\right)\right]=0$, the imaginary part of the refraction index is neglected in the calculation of the photon emission process even in the case of large $\chi_\gamma$.
Similarly, in the following we consider the effect of the real part of the index of 
refraction $ \Delta 
{n} \left( 
\chi_{\gamma}\right)=\Re \left[ \Delta\tilde{n} \left( 
\chi_{\gamma}\right)\right] $ in the calculation of photon emission distribution as we 
are interested in SCC photon spectrum and its difference from the Compton one.
For further considerations related to electron positron pair production in a 
polarized vacuum, we note that backreactions from pair creation, damping of an 
incident photon due to the decay into electron-positron pair and the radiative 
energy loss of a charged particle should be self-consistently taken into 
account \cite{Yatabe2018,Grishin2022}.
Thus the exact treatment of the interaction dynamics requires the development 
of a corresponding  numerical model that is beyond the scope of this paper.

Since $ \Delta n \left( \chi_{\gamma}\right)\neq0  $, the 
phase 
velocity of light is 
either higher ($ \Delta n \left( \chi_{\gamma}\right)>0  $) or lower ($ \Delta 
n \left( \chi_{\gamma}\right)<0  $) than the free space value.
While $ \Delta n $ is positive for $ \chi_{\gamma}\ll1 $ and has a maximum at $ 
\chi_\gamma\approx0.75 $, for $ 
\chi_{\gamma}\gtrsim 15 $ the value of $ \Delta n $ becomes negative, and thus 
the Cherenkov radiation vanishes \cite{Ritus1970}.
Due to the difference of a non-linear vacuum refraction index from 
unity, the 
probability of 
photon emission by a relativistic charged particle propagating in a strong EM 
field is modified.
In general, the emission of a photon in a medium is governed by 
the 
SCC process that depends on both positive and negative 
values of $ \Delta n(\chi_{\gamma}) $.
Near $ n\beta\approx1 $, where $\beta=\lvert \boldsymbol{\beta} \rvert$, 
there 
is a transition region that can be divided into 
two branches: Compton branch for $ n\beta<1 
$ and Cherenkov branch for $ n\beta>1 $.
The pure Compton and Cherenkov radiation processes represent the two limit 
cases of 
SCC 
emission.
For $ \Delta n = 0 $, the Compton branch of SCC 
radiation reduces to the non-linear Compton scattering given by 
Eq.~\eqref{W_Blackburn}.
To exceed the threshold for the pure Cherenkov radiation in the case when 
quantum theory 
is 
considered, 
the required 
Lorentz factor of 
the particle needs to be 
\begin{equation}\label{GammaCh}
	\gamma_{\mathrm{Ch}}>\dfrac{1}{\sqrt{2\Delta n 
			\left(1-\frac{\hbar\omega}{\mathcal{E}} \right) }},
\end{equation}
provided $ \Delta n > 0 $ \cite{Ginzburg1996}.
This condition corresponds to  $ F<0 $ for $ \theta=0 $ in Eq.~\eqref{F}.
In such a case, the Cherenkov branch becomes significant in the SCC 
process 
and the Cherenkov photons are emitted along the particle propagation direction 
within the Cherenkov angle $ \cos\theta_{\mathrm{Ch}}=1/\left( n\beta\right)  $
\cite{Ginzburg1996}.

However, even if the threshold for pure Cherenkov radiation is not met, 
the SCC radiation can still 
significantly differ from the traditionally considered Compton one  
\cite{Rynne1978,Rynne1979,Bonin1986}, which might be of interest for future 
particle-field 
interaction 
experiments studying the fundamental properties of a quantum vacuum.
\begin{figure*}
\centering
\includegraphics[width=1.0\linewidth]{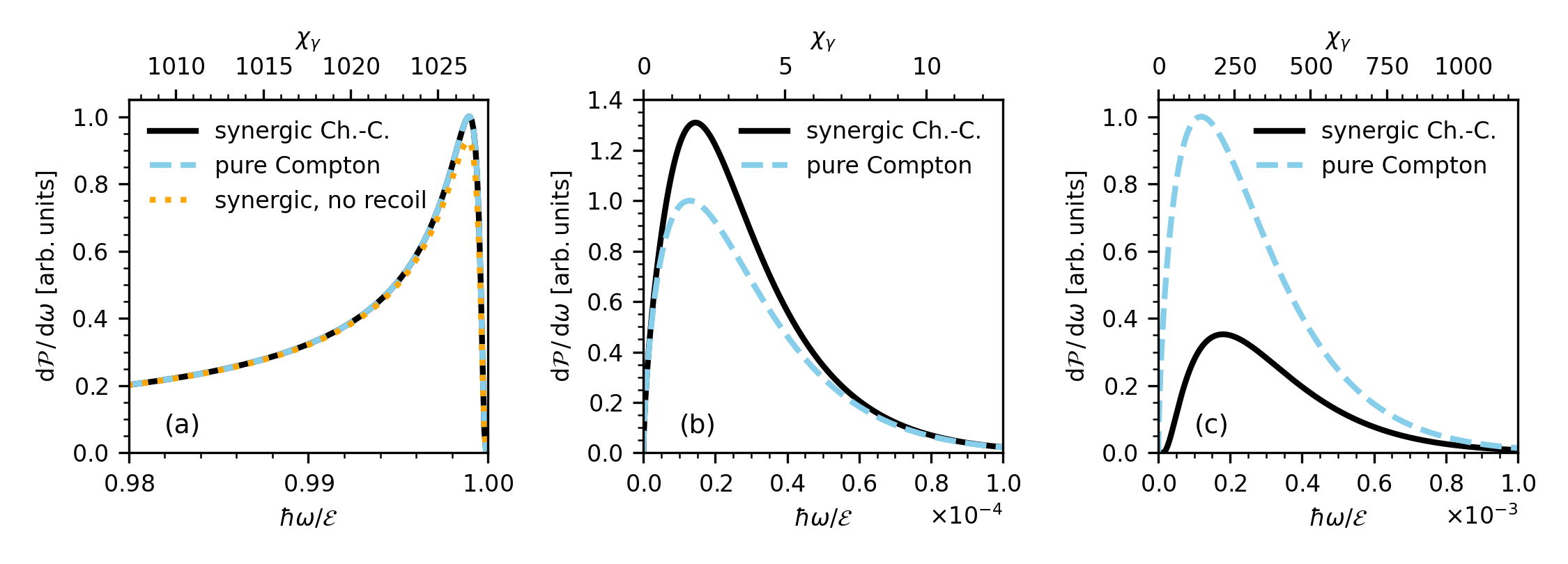}
\caption{Power of radiation $ \mathrm{d}\mathcal{P}/\mathrm{d}\omega $, 
given 
by 
Eq.~\eqref{dP}, emitted by the (a) 
40~GeV 
electron, (b) 700~GeV and (c) 6.5~TeV protons in the forward direction 
during a head-on 
collision with a strong electromagnetic field of intensity  
(a) $10^{25}~\mathrm{W/cm^{2}} $ and (b,c)
$5\times10^{26}~\mathrm{W/cm^{2}} $ in the Cherenkov-Compton 
process that accounts for the effect of vacuum polarization, i.e. $\Delta n 
\neq 0$. 
Solid 
(dotted) line corresponds 
to 
the 
case 
when a recoil is 
(is 
not) considered in Eq.~\eqref{F}, respectively. The blue dashed curve 
shows the power emitted in the pure Compton process in which vacuum 
polarization is neglected, i.e. $ \Delta 
n=0 $.}
\label{fig:fig01}
\end{figure*}
In the following, we consider the interaction of charged particles (electron, 
proton) with a strong EM field that will be provided by the 
next generation of lasers.
Using Eq.~\eqref{dP} we have calculated the emission spectra of the SCC process 
for
respective cases, considering for clearness $ 
\Delta 
n(\chi_{\gamma}) $ averaged over both polarizations.
Under weak-field approximation, the presented approach can also describe the 
synchrotron-Cherenkov radiation in a strong constant magnetic field 
\cite{Toll1952,Schwinger1976}. Such an interaction setup is relevant for 
studying the photon emission in the strong field of magnetars \cite{Duncan1992} 
or near black holes \cite{Hirotani2016}.

At first, we consider the interaction of a $  40~\mathrm{GeV} $ electron beam with a 
laser field of intensity $10^{25}~\mathrm{W/cm^{2}}$, thus the electrons reach 
$ \chi\approx10^{3} $.
For achieving such a high laser intensity with multi-PW lasers 
\cite{Danson2019,ELIwhitebook,Weber2017,Turner2022,Zeus2021}, one could 
consider the concept 
of 
relativistic flying mirrors 
\cite{Bulanov2003,Naumova2004,Bulanov2013,Vincenti2019}.
For $ \chi\gg1 $, a sharp peak close to the initial electron energy 
appears 
in the Compton emission spectrum, which means that photons with $ 
\hbar\omega\approx\mathcal{E} $ are predominantly radiated 
\cite{Esberg2009,Tamburini2021}.
Since for $ \chi_{\gamma}\gg1 $ we get $ \Delta n <0 $ from Eq.~\eqref{DeltaN}, 
the 
photon emission spectra should be suppressed near the spectral tip.
In Fig.~\ref{fig:fig01}(a) we show the emission spectra for the following 
processes: pure Compton (dashed line), SCC (solid line)
and SCC with neglected recoil term $1-\hbar\omega/\mathcal{E}$ in Eq.~\eqref{F} 
(dotted line).
As shown, there is no visible difference between the emission 
spectra in case of pure Compton and SCC radiation.
While the recoil term $ 1-\hbar\omega/\mathcal{E} $ can be considered 
insignificant for quantum Cherenkov radiation \cite{Ginzburg1996}, it is shown 
that in the more general case of Cherenkov-Compton radiation, this 
term becomes crucial as it controls the amount of energy emitted due to the 
effects of vacuum polarization.
For this particular case its neglecting results in obtaining erroneous results 
that underestimate the emitted power by approximately 10\% near the spectral 
tip in the case of SCC.
We see that the recoil effect thus plays an important role in SCC 
radiation as it considerably reduces the effect of vacuum polarization on 
photon emission spectra near $ 
\hbar\omega\approx\mathcal{E} $, see Eq.~\eqref{F}.
As a result, in the case of the interaction with an ultra-relativistic 
electron characterized by $ \chi\gg1 $, the evidence of photon spectra 
modification caused by vacuum 
polarization in a strong EM field is dramatically suppressed just due to the 
recoil.

For direct observation of the Cherenkov effect it is reasonable to 
consider photons 
characterized by $ \chi_{\gamma}\approx0.75 $, as in this case $ \Delta n 
$ reaches the maximum value while electron-positron pair production is 
considerably diminished.
Since $ \Delta n $ acquires a positive constant value for $ 
\chi_{\gamma}\ll1 
$, (see Eq.~\eqref{DeltaN}), it is natural to study the Cherenkov branch of SCC radiation of an 
electron 
in 
this limit 
\cite{Dremin2002,Bulanov2019,Macleod2019,Jirka2021,Sasorov2022}.
We note that to exceed the threshold for pure Cherenkov radiation given 
by 
Eq.~\eqref{GammaCh}, the 
Lorentz factor 
$ \gamma\gg1 $ is needed.
In such a case, photons with $ \chi_{\gamma}\gg1 $ are predominantly 
emitted while Cherenkov photons are only emitted at the low energy end of the 
spectra ($ \chi_{\gamma}\ll1 $).
The detailed analysis of photon emission in this case presented in 
Ref.~\cite{Sasorov2022} shows that the Cherenkov cone is not distinctly 
evident for the studied range of parameters.

In contrast to the case of emitting electrons, the Cherenkov branch tends 
to dominate for heavier particles
\cite{Erber1976}.
As an example we consider the interaction of a 700 GeV proton, whose energy
is an order of magnitude below the current record 
\cite{Cern2016}, with the EM field of intensity 
$5\times10^{26}~\mathrm{W/cm^{2}}$.
As shown 
in Fig.~\ref{fig:fig01}(b), the SCC process (solid line) enhances the 
peak of photon emission spectra by 30\% in comparison to pure Compton 
radiation (dashed line)
in which vacuum polarization is neglected.
When vacuum polarization is 
taken into account, by 20\% more energy is emitted in the form of 
photons 
with $ 
1\lesssim\chi_{\gamma}\lesssim8 $ that, in turn, enables more efficient pair 
production.
We note, that in the collision of a high-energy proton beam and a strong laser field, merging of laser photons can occur due to the polarization of vacuum \cite{DiPiazza2008}.

The Compton-Cherenkov process can also significantly suppress the emission of 
photons with $ \chi_{\gamma}\gg 1 $.
We note that in galactic centers protons can be accelerated up to PeV range 
\cite{Hess2016}, however,
to demonstrate this phenomenon in terrestrial laboratory, we consider a proton 
of today's available 
energy 
6.5~TeV
\cite{Cern2016}.
In such as case, photons with $ \chi_{\gamma} \gg1$ are emitted.
Since $ \Delta n(\chi_{\gamma}\gg1) <0 $, the SCC process causes the 
quenching 
of photon emission.
In Fig.~\ref{fig:fig01}(c) we show that in such a case, the peak of photon 
emission spectra is reduced by 65\% and the total emitted energy is lower by 
more than 50\% compared to pure Compton scattering.
We note that for the presented parameters, the protons are characterized by 
$ \chi\approx10^{-5}$ and $10^{-4} $, respectively, and the formation length 
for photon emission is 
on the 
order of $ \lambda/10 $, where wavelength $ \lambda=1~\mathrm{\mu m} $ is 
assumed.

We calculated photon energy distribution in Compton-Cherenkov process describing the emission of a charged particle in a vacuum 
polarized by a very strong electromagnetic field.
This phenomenon is important for the fundamental QED processes present in 
strong fields near magnetars and black holes in space and can be studied in 
collisions of accelerated particles with a strong field produced by the new 
generation of lasers.
While for electrons characterized by $ \chi\gg1 $ the effect of 
vacuum 
polarization on emission spectra peak is not expected just due to the 
recoil, for today's available 6.5~TeV (700~GeV) protons accelerated with 
the LHC accelerator \cite{Cern2016}, 
the synergic 
nature of photon emission leads to suppression (enhancement) of emitted energy by tens of percent compared to a pure Compton process (i.e. when vacuum polarization 
effects are neglected).
The features of enhancement and reduction of high-energy photon 
emission 
in 
the Compton-Cherenkov 
process can be exploited as a 
signature of vacuum texture in laser-particle experiments.

We would like to thank Antonino Di Piazza, Rashid Shaisultanov and 
Alexander John MacLeod for 
useful 
discussions.
Supported by the project Advanced research using high intensity laser produced photons and particles (ADONIS) CZ.02.1.01/0.0/0.0/16\_019/0000789 from European Regional Development Fund (ERDF).

\bibliography{scc}{}
\bibliographystyle{apsrev4-2}
\end{document}